\documentclass[a4paper,11pt,notitlepage]{article}
\usepackage[top=2cm,bottom=2cm,left=2cm,right=2cm,includefoot]{geometry}
\usepackage[margin=10pt,font=small,
            labelfont=bf,
            labelsep=quad,figurename=Fig.,
	    tablename=Tab.]{caption}
\usepackage{multirow}

\usepackage{setspace}
\usepackage{indentfirst}
\usepackage{array}
\usepackage{threeparttable}

\usepackage{bm}
\usepackage{amsmath}
\usepackage{amsbsy}
\usepackage{amstext}
\usepackage{amsfonts}

\usepackage[numbers,super,sort&compress]{natbib}
\usepackage{mciteplus}

\usepackage{graphicx}
\usepackage[rgb]{xcolor}

\usepackage{hyperref}
\hypersetup{colorlinks,citecolor=red,linkcolor=blue}

\usepackage[pagewise]{lineno}

\usepackage[T1]{fontenc} 

\usepackage{times} 
\usepackage{mathptmx} 

\usepackage{multirow} 

\begin{document}

\title{Investigations of the Underlying Mechanisms of HIF-1$\alpha$ and CITED2 Binding to TAZ1}

\date{\today}

\maketitle
\centerline{Wen-Ting Chu$^{1}$, Xiakun Chu$^{2}$, Jin Wang$^{1,2*}$}
\begin{center}	
	{\small	
	$^1$\emph{State Key Laboratory of Electroanalytical Chemistry, Changchun Institute of Applied Chemistry, Chinese Academy of Sciences, Changchun, Jilin, 130022, China\\}	
	$^2$\emph{Department of Chemistry \& Physics, State University of New York at Stony Brook, Stony Brook, NY, 11794, USA\\}
	$^*$\emph{Corresponding Author: jin.wang.1@stonybrook.edu\\}}
\end{center}

\newpage 
 
\begin{abstract}
The TAZ1 domain of CREB binding protein is crucial for transcriptional regulation and recognizes multiple targets.
The interactions between TAZ1 and its specific targets are related to the cellular hypoxic negative feedback regulation.
Previous experiments reported that one of the TAZ1 targets CITED2 is an efficient competitor of another target HIF-1$\alpha$.
Here by developing the structure-based models of TAZ1 complexes
we have uncovered the underlying mechanisms of the competitions between HIF-1$\alpha$ and CITED2 binding to TAZ1.
Our results are consistent with the experimental hypothesis on the competition mechanisms and the apparent affinity.
In addition, the simulations prove the dominant position of forming TAZ1-CITED2 complex in both thermodynamics and
 kinetics.
For thermodynamics, TAZ1-CITED2 is the lowest basin located on the free energy surface of binding in the ternary system.
For kinetics, the results suggest that CITED2 binds to TAZ1 faster than HIF-1$\alpha$.
Besides, the analysis of contact map and $\phi$ values in this study will be helpful for further experiments on TAZ1 systems.
\end{abstract}

\section{Introduction}
Intrinsically disordered proteins (IDPs) \cite{wright1999intrinsically,dunker2001intrinsically,uversky2002natively} 
behave as disordered/unstructured forms at physiological conditions in isolated states,
but sometimes undergo conformational changes to folded form upon binding to the partners \cite{dyson2002coupling,dyson2005intrinsically}.
Such binding-coupled-folding scenario has significantly refreshed our understanding on the protein structure-function paradigm.
Generally, IDPs are widely involved in many critical physiological processes, 
such as transcription and translation regulation, cellular signal transduction, protein phosphorylation, and molecular assembles \cite{gavin2002functional,ho2002systematic}.

\begin{figure}
\centering
\includegraphics[width=0.8\textwidth]{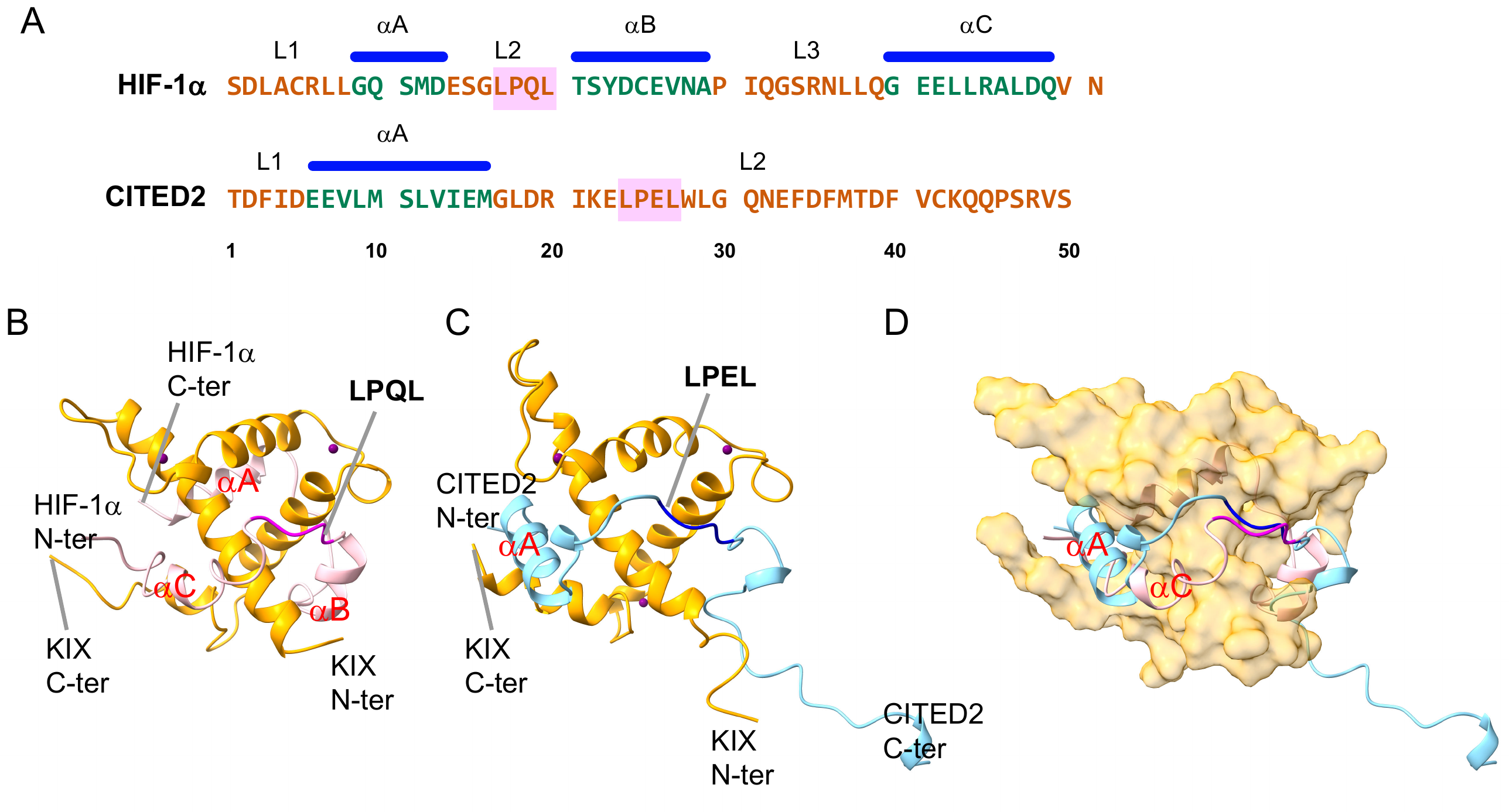}
\caption{
The sequences (A) and binding postures (B-D) of HIF-1$\alpha$ (51 a.a., corresponding to 776-826 in 1L8C) and CITED2 (50 a.a., corresponding to 220-269 in 1R8U).
HIF-1$\alpha$ includes 3 main alpha helices $\alpha$A (783-788), $\alpha$B (796-804), $\alpha$C (815-824), and 3 main loop regions, L1 (776-782), L2 (789-795), L3 (805-814).
LPQL motif is included in L2 region.
CITED2 has only one helix, $\alpha$A (225-235), and 2 main loop regions, L1 (220-224), L2 (236-269).
LPEL motif is included in L2 region.
The complex structures (B, TAZ1-HIF-1$\alpha$; C, TAZ1-CITED2) are extracted from the NMR structures 1L8C and 1R8U.
Their superimposed structure is illustrated in the panel D.
$\alpha$A, LPQL of HIF-1$\alpha$ and $\alpha$, LPEL of CITED2 share the same binding surface of TAZ1.
The alpha helices of HIF-1$\alpha$ and CITED2 are labeled both in the sequences and in the complex figures.
The essential conserved motifs are emphasized with pink boxes (in the sequences) and with magenta and dark blue cartoons (in the complex figures).
}
\label{fig:seq}
\end{figure}

The transcriptional adaptor zinc-binding 1 (TAZ1) is a protein domain of CREB binding protein (CBP),
which plays an important role in the transcriptional regulation \cite{wright2015intrinsically}.
One of its binding partners, the $\alpha$-subunit of the transcription factor (hypoxia inducible
factor) HIF-1 (HIF-1$\alpha$),
interacts with TAZ1 through its intrinsically disordered C-terminal transactivation domain,
which is related to the transcriptional regulation of genes that are crucial for cell survival during low levels of oxygen
\cite{arany1996essential,dames2002structural,freedman2002structural}.
Another binding partner CITED2, which occupies a different but partially overlapped binding site from that of HIF-1$\alpha$,
acts as a negative feedback regulator that attenuates HIF-1 transcriptional activity by competing for TAZ1 binding through its own disordered transactivation domain
\cite{bhattacharya1999functional,de2004interaction,freedman2003structural}.
Intriguingly, both the two ligands of TAZ1 have a conserved LP(Q/E)L region (LPQL in HIF-1$\alpha$; LPEL in CITED2)
that is essential for negative feedback regulation \cite{dames2002structural,freedman2002structural,de2004interaction,freedman2003structural}.
These LPQL and LPEL domains bind with the same place of TAZ1 surface at bound state (see \autoref{fig:seq}).

Though the structures of both TAZ1-HIF-1$\alpha$ and TAZ1-CITED2 complexes have been deposited to Protein Data Bank \cite{dames2002structural,de2004interaction},
little is known about how binding affinity or binding mechanism will be influenced if two ligands co-exist.
Both HIF-1$\alpha$ and CITED2 bind to TAZ1 with the same high affinity ($K_d=10\pm2$ nM) \cite{berlow2017hypersensitive}.
In addition, HIF-1$\alpha$ and CITED2 utilize partially overlapped binding sites to form complexes with TAZ1 \cite{dames2002structural,freedman2002structural,de2004interaction,freedman2003structural},
suggesting that HIF-1$\alpha$ and CITED2 are binding competitors to TAZ1.
The NMR experiments of Wright \textit{et al.} \cite{berlow2017hypersensitive} observed that TAZ1-CITED2 complex is dominant in the TAZ1:HIF-1$\alpha$:CITED2 solvation with 1:1:1 molar ratio.
The fluorescence anisotropy competition experiments found that CITED2 exhibits an apparent $K_d$ of 0.2$\pm$0.1 nM to TAZ1-HIF-1$\alpha$ complex,
while HIF-1$\alpha$ displaces TAZ1-bound CITED2 with a much higher apparent $K_d$ (0.9$\pm$0.1 $\mu$M) \cite{berlow2017hypersensitive}. 
These experimental results indicate that CITED2 is extremely effective in displacing HIF-1$\alpha$ from the TAZ1–HIF-1$\alpha$ complex.

The possible mechanism for displacement of HIF-1$\alpha$ from its complex with TAZ1 by CITED2 \cite{berlow2017hypersensitive} was proposed
that CITED2 binds to TAZ1-HIF-1$\alpha$ complex through its N-terminal region,
displacing the dynamical and weakly interacting $\alpha$ helix of HIF-1$\alpha$,
then competing through an intramolecular process for binding to the LP(Q/E)L site.
However, it is challenging to prove such replacing mechanism experimentally.
In our previous works, we have explored the complex association processes and uncovered the binding and folding mechanism as well as the key interactions with structure-based model and molecular dynamics (MD) simulations
\cite{chu2017binding,chu2012importance,chu2014dynamic,wang2012exploration,chu2017role,chu2013quantifying,chu2018quantifying}.
Here in this study, we developed the binary models of both TAZ1-HIF-1$\alpha$ binding and TAZ1-CITED2 binding as well as the ternary model of TAZ1-HIF-1$\alpha$-CITED2 binding
to unveil the mechanism of the replacing processes
(including HIF-1$\alpha$ replacing the TAZ1-bound CITED2 as well as CITED2 replacing the TAZ1-bound HIF-1$\alpha$).
Our studies quantified the free energy surface of the overall competition processes and further suggest that it is easier to form TAZ1-CITED2 complex than TAZ1-HIF-1$\alpha$ complex in thermodynamics or kinetics,
in line with the experimental findings.
This study has potential implication in for the underlying details and mechanisms of transcription regulations by TAZ1. 

\section{Results and Discussion}
\subsection{Binding free energy surface of the binary association processes}
We used a weighted structure-based model of TAZ1-HIF-1$\alpha$ and TAZ1-CITED2 complexes according to the NMR structures
1L8C \cite{dames2002structural} and 1R8U \cite{de2004interaction}, respectively.
The contact map of the weighted structure-based model was collected based on the 20 configurations in PDB, taking into account the NMR structural flexibility.
The parameters of the model were calibrated carefully in order to be consistent with the experimental measurements.
In details, the strengths of intra-chain interactions of HIF-1$\alpha$ and CITED2 were tuned according to the experimental helical content at unbound state.
It was reported that both HIF-1$\alpha$ (776-826) and CITED2 (220-269) behaved as random coil at unbound state \cite{dames2002structural,de2004interaction},
thus the helical contents of isolated HIF-1$\alpha$ and CITED2 were set to be below 10\% in our simulations.
Then the strengths of inter-chain interactions between TAZ1 and HIF-1$\alpha$ as well as between TAZ1 and CITED2 were adapted
according to the experimental dissociation constant ($K_d$ about 10 nM for both TAZ1-HIF-1$\alpha$ and TAZ1-CITED2 \cite{berlow2017hypersensitive}).
In our model, both TAZ1-HIF-1$\alpha$ and TAZ1-CITED2 complexes with inter-molecular interaction strengths of 1.10 and 0.95 have similar affinities with the experiments (corresponding to about -6.06 kT binding energy).
The experimental $K_d$ and the simulated $K_d$ of single ligand binding to TAZ1 (binary system) process are listed in \autoref{tab:kd}.

After performing replica exchange molecular dynamics (REMD) simulations for sufficient sampling with 28 replicas ranging from about 0.50 to 1.86 (simulation temperature, room/experimental temperature is about 0.99) on both TAZ1-HIF-1$\alpha$ and TAZ1-CITED2 complexes,
The weighted histogram analysis method (WHAM) algorithm \cite{kumar1992weighted,kumar1995multidimensional} was applied on the REMD trajectories to collect statistics 
and to obtain the free energy distributions as well as other characteristics at the experimental temperature.
Firstly the binding free energy profile was quantified by projecting the free energy onto the fraction of inter-molecular native contacts ($Q_{inter}$),
which can be considered as the reaction coordinate of binding.
As shown in Fig. S1, TAZ1-HIF-1$\alpha$ and TAZ1-CITED2 have similar binding affinity ($\Delta G_{bind}$), which agrees with the experimental measured $K_d$.
However, the binding barrier ($\Delta G^\ddagger$) of TAZ1-HIF-1$\alpha$ is significantly higher than that of TAZ1-CITED2 
($\Delta G^\ddagger$ (TAZ1-HIF-1$\alpha$) is about two times of $\Delta G^\ddagger$ (TAZ1-CITED2)).
The thermodynamic results suggest that when binding to TAZ1, the kinetic binding rate of HIF-1$\alpha$ should be lower than that of CITED2, though the similar binding stabilities are established.

The flexible binding or binding coupled folding behavior can be illustrated by the free energy surface along $Q_{inter}$, $Q_{intra}$ and along $Q_{inter}$, helical content (Fig. S2).
$Q_{intra}$ is the fraction of intra-molecular native contacts, which acts as folding reaction coordinates.
Upon binding, $Q_{intra}$ and the helical content of HIF-1$\alpha$ change from 0.30 to 0.75 and 0\% to 35\% (bottom of the basins of unbound and bound states), respectively.
But the changes of the $Q_{intra}$ and helical content of CITED2 during binding are not as remarkably large as that of HIF-1$\alpha$,
they alter from 0.24 to 0.42 and from 0\% to 17\%, respectively.
In addition, it is obvious that at the transition states, the $Q_{intra}$ and helical content of HIF-1$\alpha$ or CITED2 are similar as that at the unbound state,
but are different from that at the bound state.
This accords with the induce-fit binding mechanism.

\begin{table}
\centering
\caption{The experimental $K_d$ (as well as apparent $K_d$), the simulated binding energy (the energy difference between the final state and the initial state, $\Delta$G) of different systems.
The experimental $K_d$ values are extracted from the ref\cite{berlow2017hypersensitive}.
Note that the experimental apparent $K_d$ does not equal to the simulated $K_d$ because of the different calculating method.
The method of obtaining the simulated $K_d$ is described in the SI Appendix Method section.}
\label{tab:kd}
\begin{tabular}{lllll}
\hline\hline
\multicolumn{4}{l}{Binary system}                                     \\
\hline
process   &  Exp. $K_d$ (nM) &  Simu. $K_d$ (nM) & Simu. $\Delta$G (kT) \\
UB to HB  &  10              &  2.515    &   -6.75             \\
UB to CB  &  10              &  2.465    &   -6.76             \\
\hline
\multicolumn{4}{l}{Ternary system}                                     \\
\hline
process   &  Exp. app. $K_d$ (nM) &  Simu. $K_d$ (nM) & Simu. $\Delta$G (kT) \\
UB to HB  &  900        &  62.8      &   -5.14           \\
UB to CB  &  0.2        &  1.38      &   -7.05           \\
\hline\hline
\end{tabular}
\end{table}

\subsection{Thermodynamic mechanism of ternary TAZ1-HIF-1$\alpha$-CITED2 complex}
\begin{figure}
\centering
\includegraphics[width=0.8\textwidth]{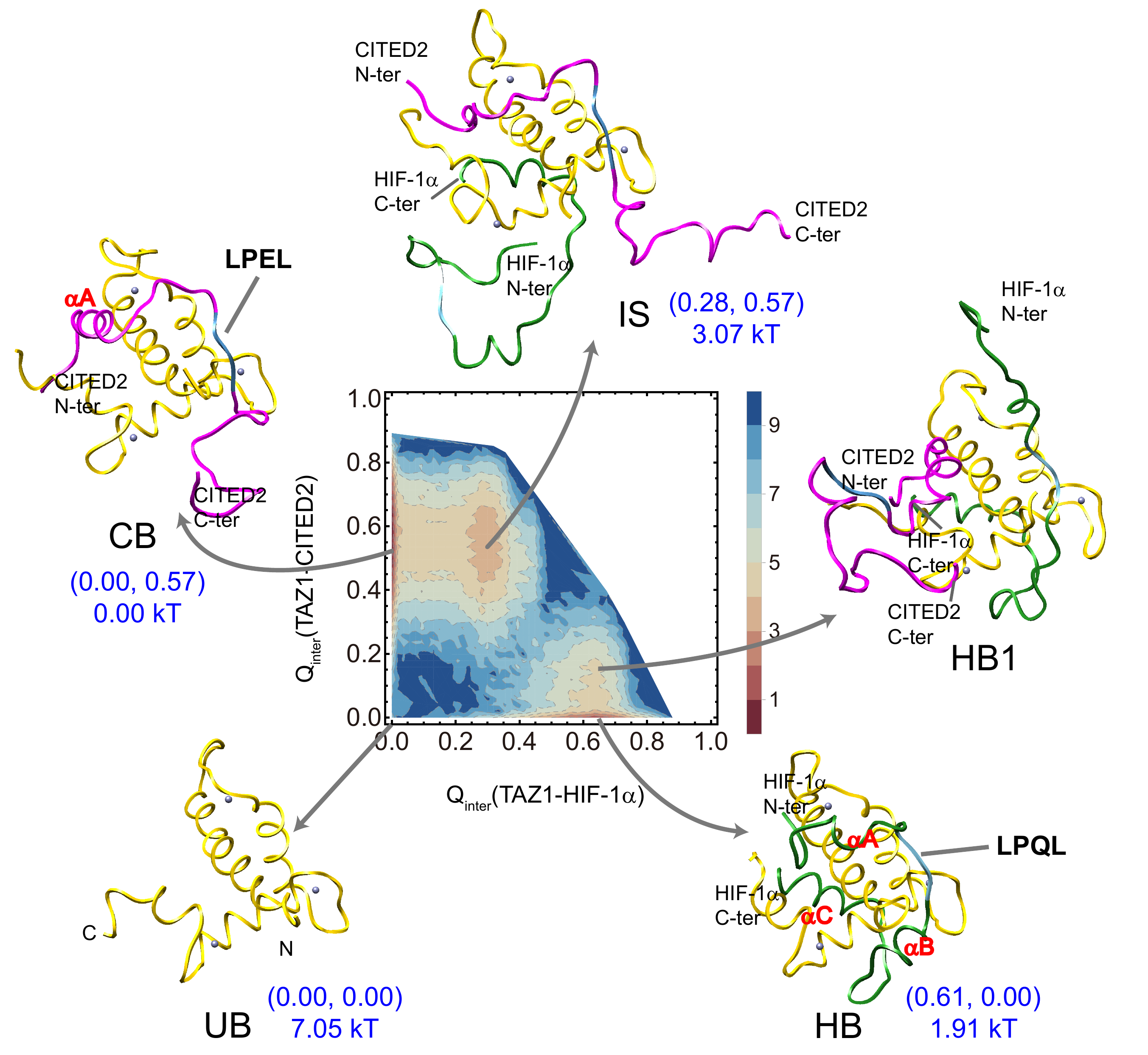}
\caption{
The free energy surface at experimental temperature as a function of $Q_{inter}$ (TAZ1-HIF-1$\alpha$) and $Q_{inter}$ (TAZ1-CITED2),
as well as the main states on the free energy surface (HB1 is not a basin on free energy surface but an area near HB state with part of CITED2 bound).
The values of the reaction coordinates (two $Q_{inter}$ values) as well as the free energy at these main states are labeled in this figure.
TAZ1 is shown in yellow ribbons (the bound Zn$^{2+}$ ions are shown with purple spheres), HIF-1$\alpha$ is shown in green ribbons, and CITED2 is shown in magenta ribbons.
The LPQL and LPEL motifs are colored in blue.
The free energy unit is kT (k is Boltzmann constant).
}
\label{fig:tri-2i}
\end{figure}

After obtained the tuned TAZ1-HIF-1$\alpha$ and TAZ1-CITED2 models, we constructed the ternary TAZ1-HIF-1$\alpha$-CITED2 model
by putting HIF-1$\alpha$ and CITED2 together with TAZ1 into the same simulation sphere.
REMD simulations with 28 replicas ranging from 0.50 to 1.86 temperature were performed on ternary TAZ1-HIF-1$\alpha$-CITED2 complex, 
with both HIF-1$\alpha$ and CITED2 unbound as the initial state.
Firstly, the free energy at experimental temperature was obtained and projected on both the binding reaction coordinates of TAZ1-HIF-1$\alpha$ and TAZ1-CITED2
($Q_{inter}$ (TAZ1-HIF-1$\alpha$) and $Q_{inter}$ (TAZ1-CITED2)).
As shown in \autoref{fig:tri-2i}, three main lower basins and other smaller basins can be recognized on the free energy surface at experimental temperature.
The lowest basin corresponds to CITED2 bound state (CB state, 0.00 kT), which means that TAZ1-CITED2 is the dominant state of all.
This is consistent with the NMR results of ref \cite{berlow2017hypersensitive}.
HIF-1$\alpha$ bound basin (HB state) is about 1.91 kT higher than CB state,
which is another bound state.
Because the LPQL region of HIF-1$\alpha$ and LPEL region of CITED2 share the same binding site on the surface of TAZ1,
HIF-1$\alpha$ and CITED2 can not fully bind with TAZ1 at the same time.
The IS state, with HIF-1$\alpha$ partly bound and LPEL of CITED2 occupied the binding site,
is the main intermediate state between HB and CB states with about 3.07 kT less stable than CB state.
The $\alpha$C helix and C-terminus of HIF-1$\alpha$ bind with one side of TAZ1; $\alpha$A helix and LPEL motif of CITED2 bind with the other side of TAZ1 at the IS state. 
The other states, including the both unbound state (UB state, about 7.05 kT higher than CB state) and 
the HB1 state near HB state (with part of HIF-1$\alpha$ bound, more than 5 kT higher than CB state),
have much higher free energy values than the CB state.
The HB1 state contains a few configurations with similar free energy values and with $\alpha$A of CITED2 bound
($Q_{inter}$ (TAZ1-HIF-1$\alpha$) about 0.61 and $Q_{inter}$ (TAZ1-CITED2) $\sim$ 0.2).
Unlike IS state, the HB1 has LPQL bound in the binding site but LPEL not bound.
These states detected on the free energy surface of TAZ1 complexed with HIF-1$\alpha$ and CITED2 are consistent with 
that proposed in the schematic mechanism for displacement of HIF-1$\alpha$ from its complex with TAZ1 by CITED2 \cite{berlow2017hypersensitive}.

In addition, it is worth noting that though TAZ1-HIF-1$\alpha$ complex and TAZ1-CITED2 complex have similar binding affinity of the binary system in both experiment and theory,
the TAZ1-HIF-1$\alpha$ state (HB) and TAZ1-CITED2 state (CB) have different free energy value in the ternary system.
The free energy results suggest that CITED2 has more opportunity to bind to TAZ1 than HIF-1$\alpha$ in the ternary system \autoref{tab:kd}.
The binding barrier data discussed above (Fig. S1) indicates that the binding rate of CITED2 associated with TAZ1 is higher than that of HIF-1$\alpha$ associated with TAZ1,
suggesting that CITED2 may bind to TAZ1 before HIF-1$\alpha$ in the ternary system.
Moreover, in the ternary system, the binding free energy of HIF-1$\alpha$ in the ternary system (-5.14 kT, see \autoref{tab:kd}) is higher than that in the binary system (-6.75 kT); 
the binding free energy of CITED2 in the ternary system (-7.05 kT) is lower than that in the binary system (-6.76 kT).
The results suggests that the binding affinity of CITED2 to TAZ1 is much higher than HIF-1$\alpha$ to TAZ1 in the ternary system, which is consistent with the tendency of experimental apparent $K_d$ (\autoref{tab:kd}).

\begin{figure}
\centering
\includegraphics[width=0.7\textwidth]{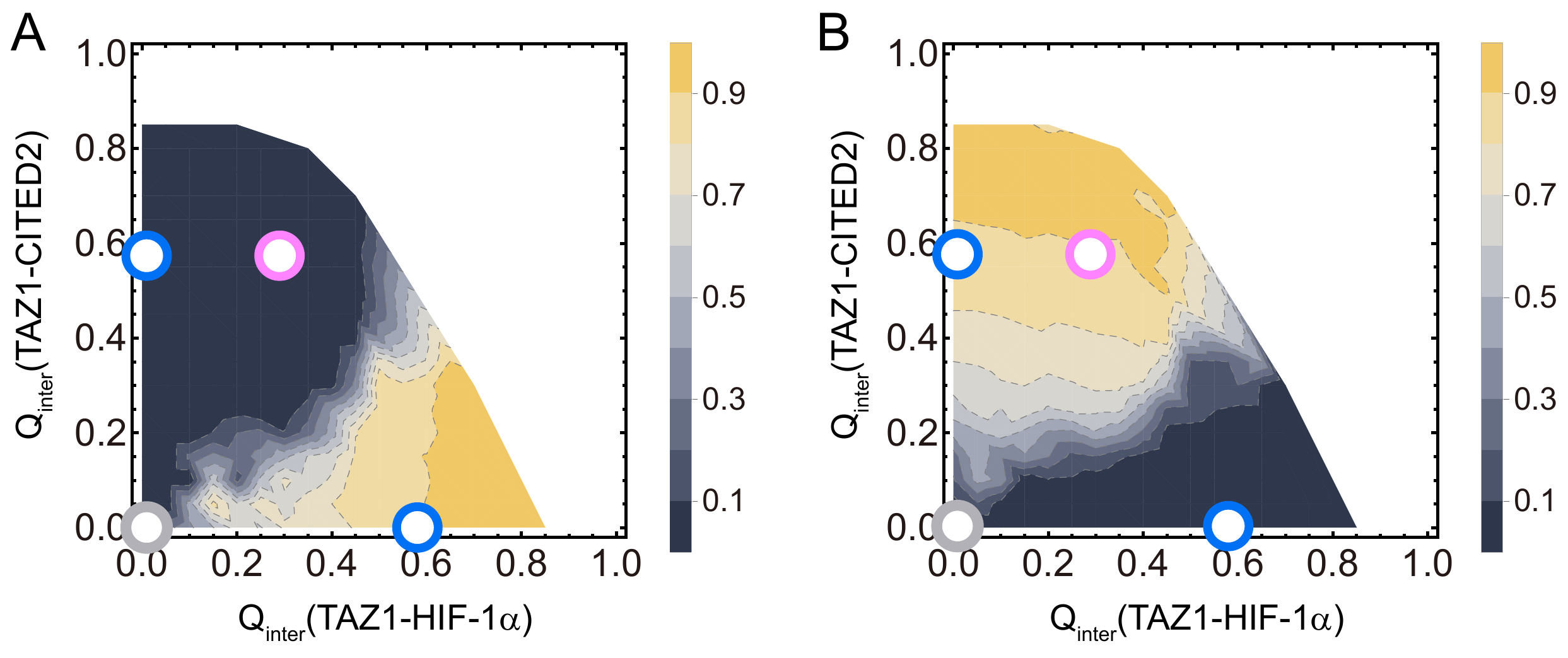}
\caption{
Mean $Q_{inter}$ between TAZ1 and LPQL of HIF-1$\alpha$ (A) and $Q_{inter}$ between TAZ1 and LPEL of CITED2 (B) 
as a function of $Q_{inter}$ (TAZ1-HIF-1$\alpha$) and $Q_{inter}$ (TAZ1-CITED2).
The locations of UB, single-ligand-bound (HB or CB), and IS states are labeled as gray, blue, and pink circles.
}
\label{fig:mean-QL}
\end{figure}

We then calculated the distribution of the helical content and $Q_{intra}$ on both binding coordinates in order to show the conformational changes of HIF-1$\alpha$ and CITED2 during binding.
As shown in Fig. S3, 
HB and HB1 states have similar $Q_{intra}$ (HIF-1$\alpha$) and helical content of HIF-1$\alpha$, which is much higher than that of CB and IS states.
Likewise, the $Q_{intra}$ (CITED2) and helical content of CITED2 decrease from CB and IS states to the HB and HB1 states.
Similar behavior can also be found in LPQL and LPEL motifs.
As shown in \autoref{fig:mean-QL}, 
LPQL/LPEL leaves from the binding site when the other ligand almost fully binds to TAZ1.
Therefore, in the direct binding process (from UB to HB/CB state), LPQL/LPEL occupies the binding site at the early part of the binding process (after transition state).
However in the replacement process (from CB to HB or from HB to CB state via IS state), 
LPQL/LPEL partly binds to TAZ1 at the late binding process (after IS state or after HB1 state).

\subsection{Kinetic mechanism of ternary TAZ1-HIF-1$\alpha$-CITED2 complex}
\begin{figure}
\centering
\includegraphics[width=0.8\textwidth]{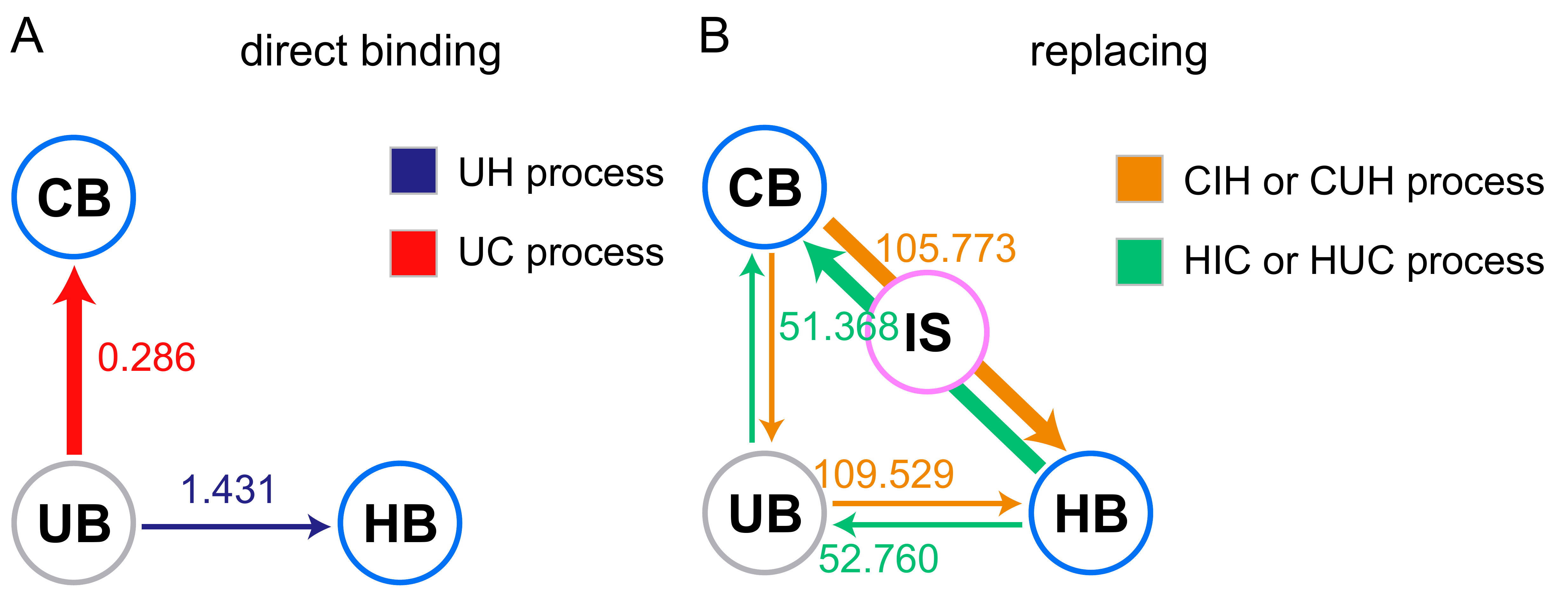}
\caption{
Direct binding (starting from UB state) and replacing (starting from HB or CB state) processes.
The probability of each binding or replacing process is roughly illustrated with the width of the arrow.
The binding time (mean $FPT_{on}$) of each process is labeled on the arrow.
}
\label{fig:scheme}
\end{figure}

As shown in the free energy profile in \autoref{fig:tri-2i}, there are four main states (UB, HB, CB, and IS) in the ternary binding system.
From the UB state, TAZ1 can bind to HIF-1$\alpha$ or CITED2 to reach HB or CB state via direct binding process (\autoref{fig:scheme}A).
HIF-1$\alpha$ can replace the TAZ1-bound CITED2 to reach HB state via IS state or UB state (CIH or CUH replacing process),
on the other hand CITED2 can replace the TAZ1-bound HIF-1$\alpha$ to reach CB state via IS state or UB state (HIC or HUC replacing process, see \autoref{fig:scheme}B).
Aiming to explore the details in the binding processes of the two ligands to TAZ1,
we performed kinetic simulations with different initial states at the experimental temperature.
Firstly, for direct binding (UB as the initial state),
mean binding time (mean first passage time, mean $FPT_{on}$) values for direct binding are 1.431 ns and 0.286 ns, respectively (\autoref{tab:ki-time}).
CITED2 binds to TAZ1 faster than HIF-1$\alpha$, which is consistent with the analyses above.
Aiming to obtain the binding probability for the two different ligands, 200 individual kinetic simulations started with varying both unbound
(both HIF-1$\alpha$ and CITED2 isolated in one sphere) configurations were performed.
This ternary system reaches CB state first (UB to CB state, denoted as UC pathway) in 177 of the 200 runs,
and reaches HB state first (UB to HB state, denoted as UH pathway) in the other 23 of the 200 runs.

We then investigated the replacing process.
200 kinetic runs started with varying TAZ1-CITED2 configurations (CB state) and 
200 kinetic runs started with varying TAZ1-HIF-1$\alpha$ configurations (HB state) were performed, respectively.
The mean $FPT_{on}$ of the replacement of CITED2 by HIF-1$\alpha$ is 163.202 ns, which is higher than that of the replacement of HIF-1$\alpha$ by CITED2 (60.758 ns as listed in \autoref{tab:ki-time}).
There are 2 different pathways of the replacing process, one is via the IS state (denoted as CIH or HIC pathway),
the other is via the UB state (denoted as CUH or HUC pathway).
The kinetic runs suggest that the probability of the pathway with IS state as processing state (CIH or HIC) is much higher
(more than 10 folds) than that of the pathway with UB state as processing state (CUH or HUC),
which means that the pathways via the IS state (CIH/HIC) are more favorable than those via the UB state (CUH/HUC).
The mean $FPT_{on}$ of of each possible pathway has been labeled in \autoref{fig:scheme}.
Note that these data were collected for the successful binding pathways,
as a result the mean $FPT_{on}$ from CB to HB (or from HB to CB) in \autoref{tab:ki-time} is higher than that of the CIH/CUH (or HIC/HUC) pathway due to including the unsuccessful binding attempts.
Moreover, it will need exceptionally more time (nearly 100 folds) for the process of replacement (CIH/CUH or HIC/HUC)
than for the direct binding process (UH or UC),
because the leaving process of the first binding ligand is time-consuming.
The barrier of the leaving process of the first binding ligand in CIH or HIC process is about 4 to 5 kT,
which is much higher than that of the direct binding process (UH or UC, about 1 to 2 kT).
Intriguingly, we found the mean $FPT_{on}$ values of the replacing pathways via IS and UB states (successful bindings) are similar (see \autoref{fig:scheme}).
The free energy surface profile (see \autoref{fig:tri-2i}) suggests that the barrier of the transition from one ligand bound state (CB or HB) 
to IS state is lower than the barrier from one ligand bound state (CB or HB) to UB state. 
As a result, it is easier to reach the IS state than to reach the UB state from the one ligand bound state. 
However, the second step of replacing process vis IS (from IS to CB or HB state) has a bit higher barrier 
than that via UB (from UB to CB or HB state).

\begin{table}
\centering
\caption{Kinetic binding time (mean first passage time of binding, mean $FPT_{on}$, ns) of different binding processes.
200 kinetic runs were performed with each starting state (UB, HB, or CB).}
\label{tab:ki-time}
\begin{tabular}{ll}
\hline\hline
\multicolumn{2}{l}{Direct binding}                                     \\
\hline
process                 & mean $FPT_{on}$    \\
UB to HB                & 1.431              \\
UB to CB                & 0.286              \\
\hline
\multicolumn{2}{l}{Replacing}                                     \\
\hline
process                 & mean $FPT_{on}$   \\
CB to HB                & 163.202            \\
HB to CB                & 60.758             \\
\hline\hline
\end{tabular}
\end{table}

\begin{figure}
\centering
\includegraphics[width=0.7\textwidth]{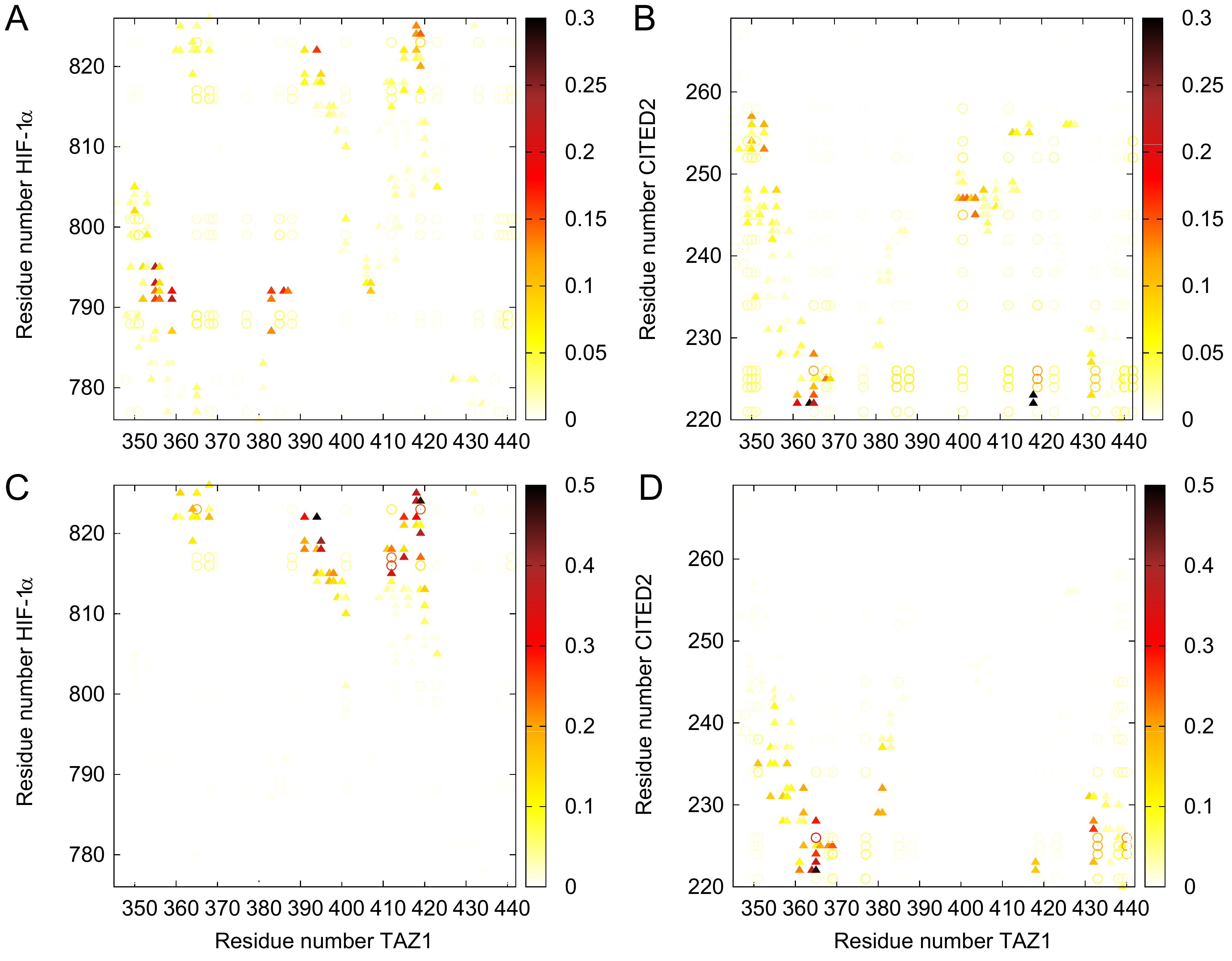}
\caption{
The probability of the interactions between TAZ1 (residue 345 to 439, as well as 3 Zn$^{2+}$) and HIF-1$\alpha$ (residue 776 to 826)
or between TAZ1 and CITED2 (residue 220 to 269) at the transition state of the binding process ($0.03<Q_{inter}<0.1$)
in the direct binding processes, UH (A) and UC (B) pathways, as well as in the replacing processes, CIH (C) and HIC (D) pathways.
Native contacts are labeled in triangles and non-native contacts are labeled as circles.
For non-native contacts, a contact is considered formed if the distance between the two residues is lower than 10.0 \AA.
}
\label{fig:contmap-all}
\end{figure}

The contact maps between TAZ1 and HIF-1$\alpha$ and between TAZ1 and CITED2 at the transition state (TS) of the different pathways are illustrated in \autoref{fig:contmap-all}
to show which part is important for the initial binding process.
Because CUH and HUC pathways include the UH and UC binding pathways as the second binding step,
here we analyze the inter-chain contact maps of UH, UC (direct), as well as CIH, HIC (replacing) pathways.
As shown in \autoref{fig:contmap-all}A and B, for the direct binding process, 
LPQL motif (residue 792-795 in 1L8C) and C-terminus of HIF-1$\alpha$ have strong interactions with TAZ1;
N-terminus and LPEL motif (residue 243-246 in 1R8U) of CITED2 are crucial for the initial binding to TAZ1.
Additionally, there are abundant non-native interactions formed in transition states, implying that TS is highly non-specific.
In contrast, for the replacement via I state (\autoref{fig:contmap-all}C and D), only C-terminus of HIF-1$\alpha$ and N-terminus of CITED2 are important for the transition state.
The interactions between LPQL/LPEL motif and TAZ1 are relatively weak or vanish.
And the non-native interactions are highly oriented. 
Therefore, the LPQL/LPEL motif may take different roles in the direct binding and replacing pathways.

\subsection{Binding order and $\phi$ value analysis}
\begin{figure}
\centering
\includegraphics[width=0.7\textwidth]{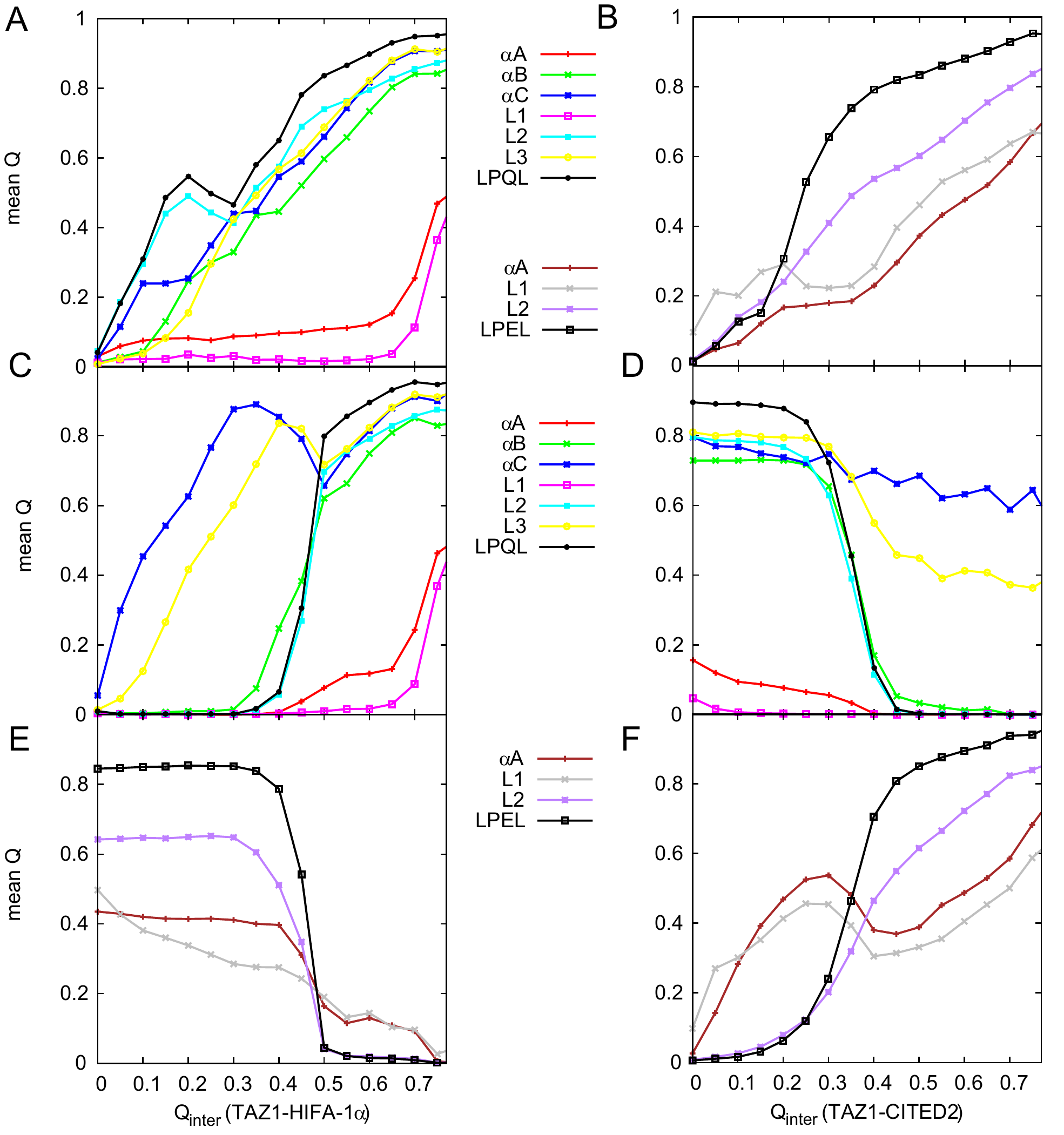}
\caption{
The mean $Q_{inter}$ between different parts of ligand (HIF-1$\alpha$ or CITED2) and TAZ1 as a function of binding 
in the UH (A), UC (B), CIH (C and E), and HIC (D and F) pathways.
Panels A, C, D show the $Q_{inter}$ curves of HIF-1$\alpha$; Panels B, E, and F show the $Q_{inter}$ curves of CITED2.
}
\label{fig:distri-all}
\end{figure}

We then divided HIF-1$\alpha$ and CITED2 into several parts for analysis (illustrated in \autoref{fig:seq})
and calculated the evolution of the interactions between these parts and TAZ1 in different pathways.
As shown in \autoref{fig:distri-all}, different pathways have different interacting regions and orders.
For HIF-1$\alpha$ direct binding processes, L2 region (including LPQL motif) and $\alpha$C reach TAZ1 first in the UH pathway (\autoref{fig:distri-all}A),
followed by the $\alpha$B helix and L3 region.
The $\alpha$A helix and L1 bind to TAZ1 last.
In addition, \autoref{fig:distri-all}A indicates that $Q_{inter}$ between the N-terminus (including $\alpha$A and L1) is lower than 0.2 at the basin of bound state ($Q_{inter}$ (TAZ1-HIF-1$\alpha$) about 0.61).
This suggests that the N-terminus of HIF-1$\alpha$ does not bind to TAZ1 closely at bound state.
We then calculated the $Q_{intra}$ within the different parts of the ligand.
It is clear in Fig. S4A that HIF-1$\alpha$ folds as binding to TAZ1, except for the part of $\alpha$A helix.
At unbound state, the folding level of $\alpha$C is higher than that of $\alpha$A and $\alpha$B.
The situation is a bit different in the CIH pathway.
As illustrated in \autoref{fig:distri-all}C, because of the existence of CITED2,
$\alpha$A, $\alpha$B, L1, L2 (including the LPQL motif) of HIF-1$\alpha$ begin to bind to TAZ1 until CITED2 is away from the binding site.
The C-terminus (including the $\alpha$C and L3) binds to TAZ1 first.
In contrast, the existence of CITED2 causes the folding of $\alpha$B of HIF-1$\alpha$ later and $\alpha$C earlier in the CIH pathway than that in the UH pathway (see Fig. S4A).
And this does not have an effect on the folding of $\alpha$A of HIF-1$\alpha$.

For the CITED2 direct binding processes, it seems that the L1 region of the N-terminus of CITED2 binds to TAZ1 first in the UC pathway (\autoref{fig:distri-all}B),
followed by the other parts.
And the CITED2 folds gradually as binding to TAZ1 (shown in Fig. S4B).
However, in the HIC pathway, $\alpha$A and L1 of CITED2 get in touch with TAZ1 first,
while the C-terminus, L2 region (including the LPEL motif), occupies the binding site until HIF-1$\alpha$ leaves the TAZ1.
The existence of HIF-1$\alpha$ causes the folding of $\alpha$A earlier in the HIC pathway than that in the UC pathway.
Moreover, the CIH pathway, the C-terminus (including the $\alpha$C and L3) of HIF-1$\alpha$ is the last part away from TAZ1,
which is also the first part that binds to TAZ1;
for HIC pathway, the N-terminus (including the $\alpha$A and L1) of CITED2 leaves TAZ1 last, which is the first part that binds to TAZ1.

Aiming to find out the crucial residues in the binding process, we calculated the $\phi$ value of different binding pathways.
For direct binding with UH pathway (see Fig. S5A and B),
most $\phi$ values of HIF-1$\alpha$ are lower than 0.2, except for the Gly791, Leu792, and Gln824.
The residues with relatively high $\phi$ values locate on the $\alpha$A, LPQL motif, and the C-terminus of HIF-1$\alpha$.
While in the CIH pathway with the existence of CITED2 (see Fig. S6A and B),
the $\phi$ values of $\alpha$A and LPQL motif decrease and the $\phi$ values of the C-terminus increase when compared with the UH pathway.
These results agree with the analysis of native contact distribution above.
For both UC and HIC pathways (see Fig. S5C and D, Fig. S6C and D),
the N-terminus of CITED2 has higher $\phi$ values than other parts of CITED2,
especially the Phe222 and Ile223.

Some experimental $\phi$ values of HIF-1$\alpha$ are labeled in Fig. S5B.
Most simulated $\phi$ values at these residue locations are lower than 0.2, in agreement with the experimental $\phi$ values.
The experimental results demonstrate that native hydrophobic binding interactions have not been created yet at the rate-limiting transition state for binding between TAZ1 and HIF-1$\alpha$, 
which is consistent with our findings that the ligand changes its conformation and folds after the transition state. 
However, the mutation V825A (red dot in Fig. S5B) has a much higher experimental $\phi$ value (0.34) than the simulated one (0.15).
We have noticed that both $\Delta \Delta G_{Eq}$ and $\Delta \Delta G_{TS}$ are negative \cite{lindstrom2018transition},
which means that this mutation will stabilize both the transition state and the bound state. 
But in the theoretical method of $\phi$ value calculation, we assume that the mutation will break the interactions between this residue and the others \cite{clementi2000topological}.
And the theoretical $\phi$ value calculation will be sensitive for the residue sites with high $\phi$ values and accurate for the $\phi$ value results driven by native contacts.
Perhaps this can explain the difference between these two $\phi$ values.

\section{Conclusions}
The TAZ1 domain of CREB binding protein is reported to be associated with two different targets that share parts of the binding site.
It has been found that in the solution of TAZ1 with two different targets HIF-1$\alpha$ and CITED2,
TAZ1 prefers to form complex with CITED2 rather than HIF-1$\alpha$.
Aiming to determine the underlying mechanism of the competitive binding between HIF-1$\alpha$ and CITED2,
we performed coarse-grained molecular dynamics simulations by developing the structure-based model of TAZ1, HIF-1$\alpha$ and CITED2 systems.
Several main states (UB, HB, CB, and IS) and a sub-state HB1 can be quantified on the free energy surface of ternary system (TAZ1, HIF-1$\alpha$, and CITED2).
The simulated mechanism is consistent with the previous experimental hypothesis about the replacing process and the apparent affinity.
The results suggest the dominant position of forming TAZ1-CITED2 complex in both thermodynamics and kinetics.
In addition, the analysis about the inter-chain contacts between TAZ1 and target shows the different binding order of each domain in direct binding and replacing processes.
In the replacing process CIH, the crucial LPQL and $\alpha$B domain of HIF-1$\alpha$ access their binding site after the intermediate state,
which is different from that in the direct binding process UH.
Besides, the different binding processes (CIH and UH) will also change the distribution of the HIF-1$\alpha$ $\phi$ values.
For the replacing process HIC, the intermediate state locates close to the CITED2 bound state.
The different binding processes (HIC and UC) will not change the distribution of the CITED2 $\phi$ values significantly.

\section{Methods}
NMR structures 1L8C \cite{dames2002structural} and 1R8U \cite{de2004interaction} were used for preparing the initial models of TAZ1-HIF-1$\alpha$ and TAZ1-CITED2 complexes.
Full-length HIF-1$\alpha$ and CITED2 are 51 and 50 a.a. proteins (776-826 and 220-269), respectively.
The initial coarse-grained C$_{\alpha}$ structure-based model (SBM) of TAZ1-HIF-1$\alpha$ and TAZ1-CITED2 complexes was
generated using SMOG on-line toolkit \cite{noel2010smog,clementi2000topological,noel2012shadow,lammert2009robustness}.
There are 3 Zn$^{2+}$ ions linked with TAZ1 with coordination bonds, modeled by one bead with 2 positive charge (+2e) for each ion.
In the present work, the weighted contact map was built with all the 20 configurations in each NMR structure.
Each native contact was identified by the CSU algorithm \cite{sobolev1999automated}.
The weighted coefficient (for intermolecular contacts and the contacts within HIF-1$\alpha$/CITED2) is the frequency of occurrence in all the configurations,
similar as the method in our previous studies \cite{chu2017binding}.
All simulations were performed with Gromacs 4.5.5 \cite{hess2008gromacs}.
The coarse grained molecular dynamics simulations (CGMD) used Langevin
equation with constant friction coefficient $\gamma =1.0$.
The cutoff length for non-bonded interactions was set to 3.0 nm.
The MD time step was set to 0.5 fs and the trajectories were saved every 2 ps.
To enhance the sampling of binding events,
a strong harmonic potential was added if the distance between the center of mass of TAZ1 and HIF-1$\alpha$, TAZ1 and CITED2
is greater than 6 nm \cite{tribello2014plumed}.
The detailed steps and settings are introduced in SI Appendix.

\section{Acknowledgments}
This work was supported by National Natural Science Foundation of China Grants (21603217, 21721003),  Ministry of Science and Technology of China Grants 2016YFA0203200.

\bibliographystyle{rsc}
\bibliography{TAZ1_abb}

\end{document}


\title{SI Appendix: \\ Investigations of the Underlying Mechanisms of HIF-1$\alpha$ and CITED2 Binding to TAZ1}

\date{\today}

\maketitle
\centerline{Wen-Ting Chu$^{1}$, Xiakun Chu$^{2}$, Jin Wang$^{1,2*}$}
\begin{center}	
	{\small	
	$^1$\emph{State Key Laboratory of Electroanalytical Chemistry, Changchun Institute of Applied Chemistry, Chinese Academy of Sciences, Changchun, Jilin, 130022, China\\}	
	$^2$\emph{Department of Chemistry \& Physics, State University of New York at Stony Brook, Stony Brook, NY, 11794, USA\\}
	$^*$\emph{Corresponding Author: jin.wang.1@stonybrook.edu\\}}
\end{center}

\newpage 

\section{Materials and Methods}
\subsection{Initial model of simulation}
NMR structures 1L8C \cite{dames2002structural} and 1R8U \cite{de2004interaction} were used for preparing the initial models of TAZ1-HIF-1$\alpha$ and TAZ1-CITED2 complexes.
Full-length HIF-1$\alpha$ and CITED2 are 51 and 50 a.a. proteins (776-826 and 220-269), respectively.
The initial coarse-grained C$_{\alpha}$ structure-based model (SBM) of TAZ1-HIF-1$\alpha$ and TAZ1-CITED2 complexes was
generated using SMOG on-line toolkit \cite{noel2010smog,clementi2000topological,noel2012shadow,lammert2009robustness}.
There are 3 Zn$^{2+}$ ions linked with TAZ1 with coordination bonds, modeled by one bead with 2 positive charge (+2e) for each ion.
In the present work, the weighted contact map was built with all the 20 configurations in each NMR structure.
Each native contact was identified by the CSU algorithm \cite{sobolev1999automated}.
The weighted coefficient (for intermolecular contacts and the contacts within HIF-1$\alpha$/CITED2) is the frequency of occurrence in all the configurations,
similar as the method in our previous studies \cite{chu2017binding}.
The potential energy function consists of both bonded and non-bonded terms.
Additionally, we introduced the charge characteristics into our SBM model to study the electrostatic interactions in this system.
As a result, the potential energy form used in this study is given in the following equation:
\begin{equation}
\begin{split}
V = {}& \sum \limits_{bonds} \epsilon _r ( r - r_0 ) ^2
+ \sum \limits _{angles} \epsilon _{h} \epsilon _\theta ( \theta - \theta_0)^2  \\
{}& + \sum \limits _{dihedral} \epsilon _{h} \epsilon _\phi ^{(n)} \left( 1 - \cos ( n \times ( \phi - \phi_0 ) ) \right)  \\
{}& + \sum \limits _{contacts} \epsilon _{ij} \left (5 \left( \frac{\sigma _{ij}}{r_{ij}} \right ) ^{12} - 6 \left( \frac{\sigma _{ij}}{r_{ij}} \right )^{10} \right) \\
{}& + \sum \limits _{non-contacts} \epsilon _{NC} \left( \frac{\sigma _{NC}}{r_{ij}} \right)^{12} + \epsilon _{DH} V _{Debye-H\ddot uckel}
\end{split}
\label{eq:SBM}
\end{equation}
In \autoref{eq:SBM}, $\epsilon _r = 100 \epsilon$, $\epsilon _\theta = 20 \epsilon$,
$\epsilon_\phi ^{(1)} = \epsilon$ and $\epsilon_\phi ^{(3)} = 0.5 \epsilon$.

\subsection*{Electrostatic interactions}
The electrostatic interactions are calculated by the $Debye-H\ddot uckel$ model,
which can quantify the strength of charge-charge attraction and repulsion at various salt
concentrations:
\begin{equation}
{V_{Debye - H\ddot uckel}} = {\Gamma _{DH}} \times {K_{coulomb}}B(\kappa )\sum\limits_{i,j} {\frac{{{q_i}{q_j}\exp \left( { - \kappa {r_{ij}}} \right)}}{{\epsilon {r_{ij}}}}}
\label{eq:DH}
\end{equation}
In \autoref{eq:DH}, $K_{coulomb} = 4 \pi \epsilon_{0} = 138.94$ $kJ \cdotp mol ^{-1} \cdotp nm \cdotp e ^{-2}$
is the electric conversion factor;
$B(\kappa )$ is the salt-dependent coefficient;
$\kappa ^{-1}$ is the Debye screening length, which is directly influenced by the
solvent ion strength (IS)/salt concentration $C_{salt}$
($\kappa \approx 3.2 \sqrt{C_{salt}} $);
$\epsilon$ is dielectric constant, which is set to 80 during the simulations.
$\Gamma _{DH}$ is the energy scaled coefficient which aims to make the total energy balanceable.
In our model, Lys and Arg have a positive point charge (+e),
Asp and Glu have a negative point charge (-e).
All the charges are placed on the $C_\alpha$ atoms.
Under the physiological ionic strengths ($C_{salt} = 0.15 M$), $\kappa$ is 1.24 $nm^{-1}$,
so we set $\Gamma _{DH} = 0.535$ in our simulations,
so that $V_{DH}$ for two opposite charged atoms located at a distance of 0.5 nm matches the native contact energy.
When a native contact is an ionic pair (salt bridge), we rescaled its interaction strength by setting $\epsilon_{DH}=0.1$
so that its energetic contribution will be comparable to other native contacts \cite{levy2007fly}.
More details of $Debye-H\ddot uckel$ model can be found in these papers
\cite{azia2009nonnative,givaty2009protein,chu2012importance,wang2012exploring}.

\subsection{Parameter calibration}
For angle and dihedral terms, some hinge regions were defined according to the structural flexibility of the NMR data,
aiming to collect the information of conformational change during ligand binding.
In this method, local interactions are weakened by decreasing the site-specific constants from the previous studies \cite{okazaki2006multiple,wang2012exploring}.
When variances of angle and dihedral are higher than 12.82 and 40.50 degrees, the potential energies of them are higher than 1.0 kJ/mol.
Because the structure of TAZ1 is stable among the NMR structures of each complex,
here we calculated the hinge regions of TAZ1 between TAZ1-HIF-1$\alpha$ and TAZ1-CITED2 to ensure the conformational flexibility of TAZ1 in the ternary complex.
Then the hinge regions of HIF-1$\alpha$ or CITED2 were measured within their structures in 1L8C or 1R8U, respectively.
In this model, if the angle or dihedral belong to the hinge regions, $\epsilon _\theta$ or $\epsilon _\phi ^{(n)}$ is rescaled by setting $\epsilon _{h}=0.01$,
mimicking the flexibility.
Otherwise $\epsilon _{h}=1$.

For the non-local attractive term (Lennard-Jones potential) in potential energy, we divided it into five parts:
intra-TAZ1, intra-HIF-1$\alpha$, intra-CITED2 terms, as well as the inter-molecular terms between TAZ1 and HIF-1$\alpha$, between TAZ1 and CITED2.
These parts have different parameters of Lennard-Jones potential ($\epsilon _{ij}$):
\begin{equation}
V_{non-bond}^{native} = \alpha_{T} V_{intra}^{TAZ1} + \alpha_{H} V_{intra}^{HIF-1\alpha} + \alpha_{C} V_{intra}^{CITED2} +
\beta_{H} V_{inter}^{TAZ1-HIF-1\alpha} + \beta_{C} V_{inter}^{TAZ1-CITED2}
\label{eq:LJ}
\end{equation}
As shown in \autoref{eq:LJ}, the $\alpha$ parameters are set for intra-molecular terms and the $\beta$ parameters are set for inter-molecular terms.
The strength $\alpha_{T}$ was set to 1.0.
Other intra-molecular parameters $\alpha_{H}$ and $\alpha_{C}$ were tuned according to the helical content of HIF-1$\alpha$ and CITED2 at unbound state.
The inter-molecular parameters $\beta_{H}$ and $\beta_{C}$ were tuned according to the dissociation constant $K_d$ in experiment.

\begin{figure}
\centering
\includegraphics[width=0.5\textwidth]{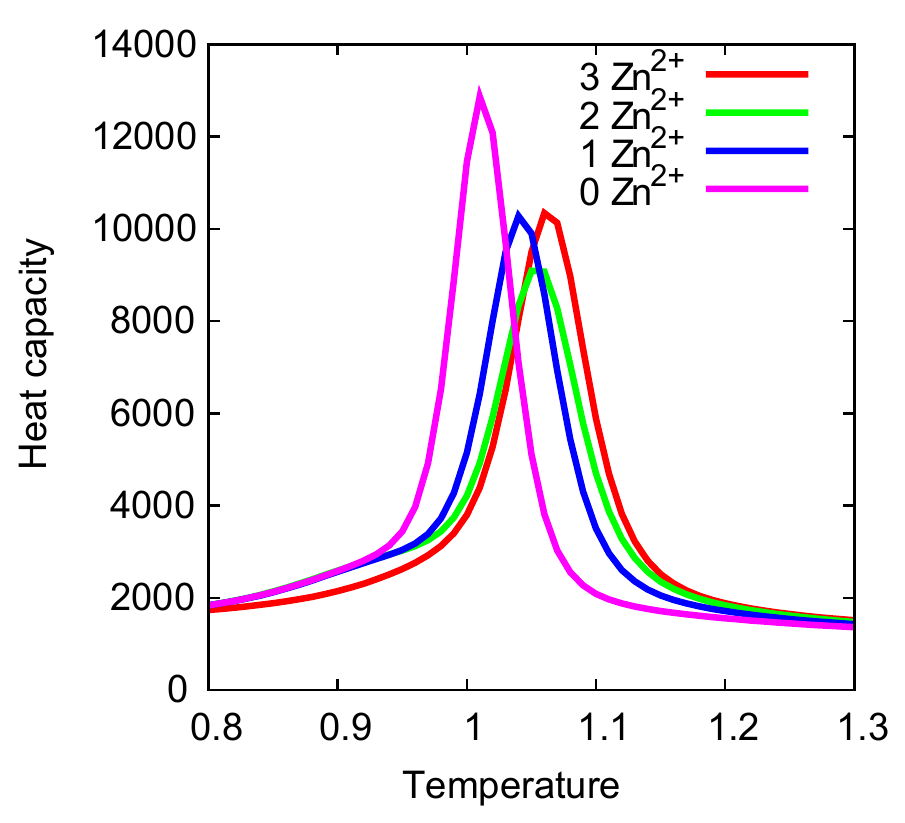}
\caption{
The heat capacity curves of TAZ1 with 3 Zn$^{2+}$ (red), 2 Zn$^{2+}$ (green), 1 Zn$^{2+}$ (blue), as well as TAZ1 without Zn$^{2+}$ (magenta).
}
\label{fig:cv}
\end{figure}

There are a little differences between the structures of TAZ1 in 1L8C and 1R8U.
Therefore we kept the intra-TAZ1 native contacts with the ratio of distances $R_{ij}$ in 1L8C and 1R8U in the range of 0.8 to 1.25,
and discarded the other native contacts to make the TAZ1 more ``flexible''.
For the MD simulation with structure-based model was run with reduced units, the simulation temperature should be calibrated firstly.
However, there is no melting temperature of TAZ1 reported.
Because there is a critical phenomenon that the three Zn$^{2+}$ ions stabilize the structure of TAZ1,
we built 4 different models of TAZ1 (TAZ1 with 3 Zn$^{2+}$, 2 Zn$^{2+}$, 1 Zn$^{2+}$, as well as TAZ1 without Zn$^{2+}$)
to find out the effect of Zn$^{2+}$ on the thermodynamic stability of TAZ1.
As shown in \autoref{fig:cv}, at temperature about 0.99, TAZ1 with 3 Zn$^{2+}$ is stable but TAZ1 without Zn$^{2+}$ unfolds.
As a result, the simulation temperature is set to 0.99, mimicking the room temperature.

In the experiment, the isolate HIF-1$\alpha$ or CITED2 was considered as "random coil"  \cite{dames2002structural,de2004interaction}.
Here we calibrated the model to make the helical content of HIF-1$\alpha$/CITED2 below 10\%.
The helical content can not decline by just altering the strength of the native contact within HIF-1$\alpha$/CITED2.
We then adjusted the dihedral potential of HIF-1$\alpha$ and CITED2 empirically by adding a term
$V(\phi) = k_\phi cos[\phi - \delta]$,
where $\delta = 297.35^{\circ}$ \cite{de2012modulation}.
We tuned the value of $k_\phi$ and found that when it equals to 1.0 or 0.9, the helical content of HIF-1$\alpha$/CITED2 is about 10\%.

\begin{figure}
\centering
\includegraphics[width=1.0\textwidth]{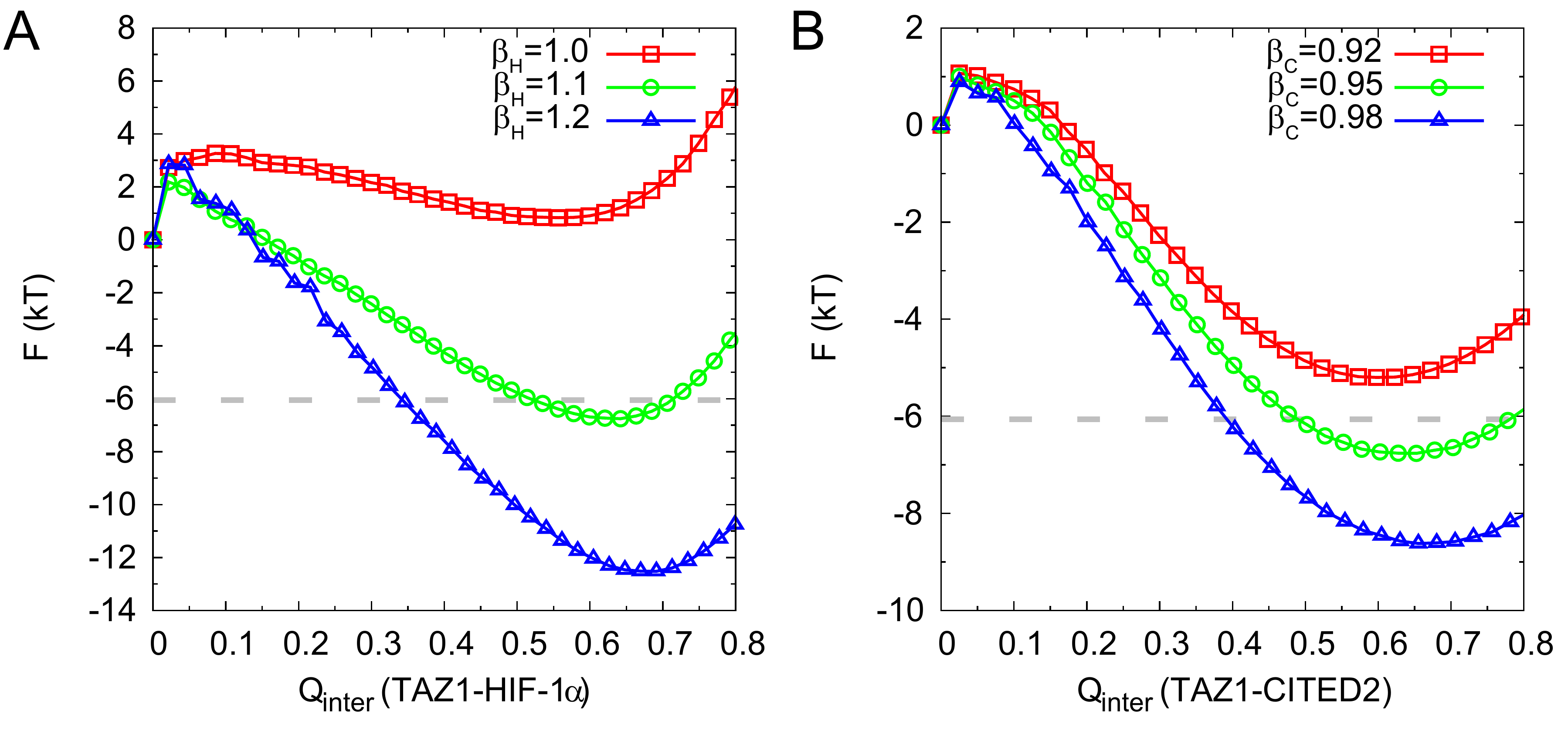}
\caption{
Free energy as a function of HIF-1$\alpha$ binding (A) and CITED2 binding (B) with different parameters. 
}
\label{fig:beta}
\end{figure}

In order to achieve sufficient sampling, TAZ1-HIF-1$\alpha$ and TAZ1-CITED2 were placed in a sphere with a radius of 6 nm,
leading to an effective concentration for the components of TAZ1 ($\left[ {C^0} \right]^{Sim}$) about 1.83 mM
(${\left[ {{C^0}} \right]^{Sim}} = \frac{{1660}}{{{V_0}}}$, 
where $V_0$ is the box volume in units of \AA$^3$, 1660 is the unit transferring constant from units of molecules per \AA$^3$ to units of mol/L, 
similar settings can be also found in previous papers \cite{law2014prepaying,wang2011multi,ganguly2011topology,wang2012exploring}).
Therefore,
\begin{equation}
{K_d} = \frac{{\left[ L \right]\left[ R \right]}}{{\left[ {LR} \right]}} = \frac{{\left( {P_{ub}{{\left[ {C^0} \right]}^{Sim}}} \right)\left( {P_{ub}{{\left[ {C^0} \right]}^{Sim}}} \right)}}{{{P_b}{{\left[ {C^0} \right]}^{Sim}}}} = \frac{{P_{ub}^2{{\left[ {C^0} \right]}^{Sim}}}}{{{P_b}}}
\label{eq:kd1}
\end{equation}
where $P_{ub}$ and $P_{b}$ are the fractions of population of unbound states and bound states at equilibrium, respectively.
From the experimental $K_d$, we can obtain the ratio of $P_{ub}/P_{b}$ at simulation condition (initial concentration $\left[ {C^0} \right]^{Sim}$). 
Then we can obtain $\Delta G$ by applying the Boltzamann distribution $\Delta G = kT\ln \left( {{P_{ub}}/{P_b}} \right)$. 
At equilibrium of the simulations, $P_{b}$ is far larger than $P_{ub}$ and close to 1. 
As a result, when the experimentally determined $K_d$ between TAZ1 and HIF-1$\alpha$/CITED2 is 10 nM \cite{berlow2017hypersensitive}, 
the binding free energy between ligand and TAZ1 in \autoref{eq:kd1} is about -6.06 kT in our model, if the effective simulation concentrations were applied. 
The method of binding free energy calculation is the same as that in the previous papers \cite{de2012modulation,law2014prepaying,wang2011multi,ganguly2011topology,chu2018quantifying}.
The strengths of inter-molecular interactions ($\beta_{H}$ and $\beta_{C}$) were tuned by performing a series of REMD simulations on TAZ1-HIF-1$\alpha$ and TAZ1-CITED2 complexes, respectively.
As shown in \autoref{fig:beta}, both $\beta_{H}$ and $\beta_{C}$ are set to be 1.1 and 0.95 in our model.

In the ternary system, we can obtain the simulated $K_d$ by using the free energy difference between bound state (HB or CB) and unbound state (UB).

\subsection{MD simulation}
All simulations were performed with Gromacs 4.5.5 \cite{hess2008gromacs}.
The coarse grained molecular dynamics simulations (CGMD) used Langevin
equation with constant friction coefficient $\gamma =1.0$.
The cutoff length for non-bonded interactions was set to 3.0 nm.
The MD time step was set to 0.5 fs and the trajectories were saved every 2 ps.
To enhance the sampling of binding events,
a strong harmonic potential was added if the distance between the center of mass of TAZ1 and HIF-1$\alpha$, TAZ1 and CITED2
is greater than 6 nm \cite{tribello2014plumed}.

\begin{figure}
\centering
\includegraphics[width=1.0\textwidth]{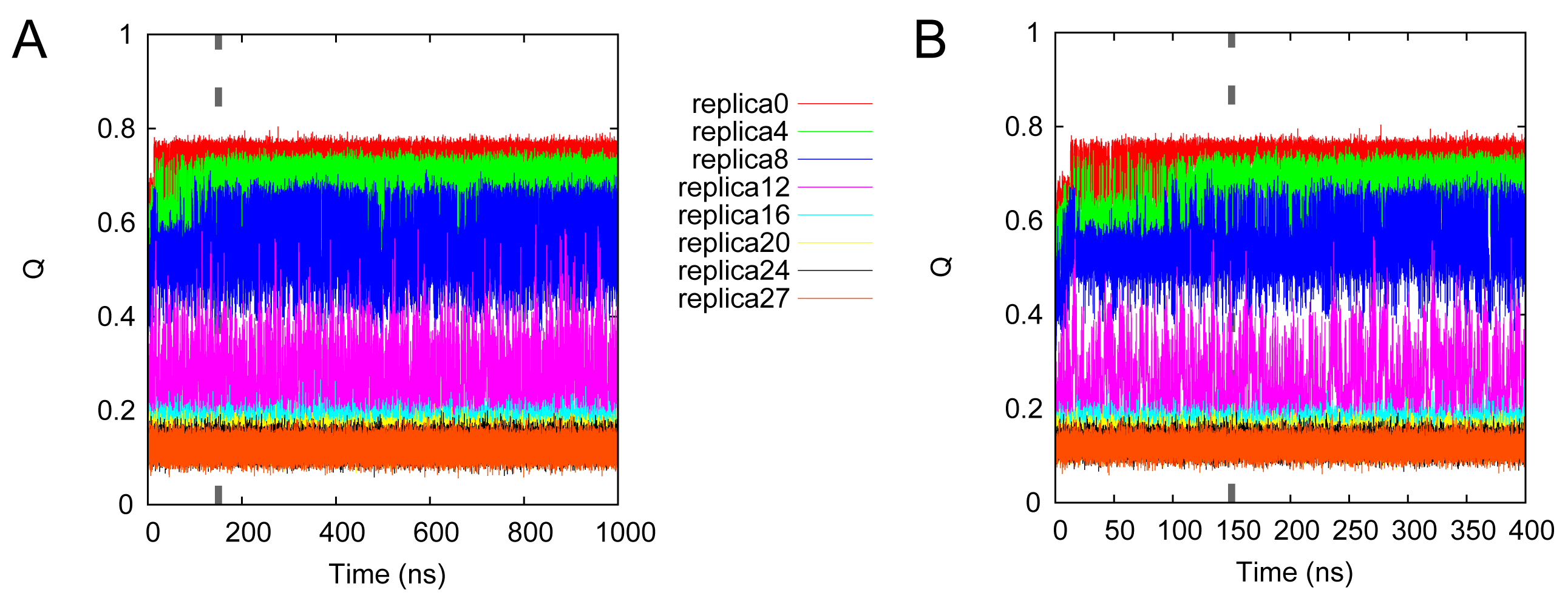}
\caption{
The fraction of all the native contacts in the ternary system as a function of simulation time of the REMD run (total 1 $\mu$s per replica).
The native contacts include the intra-molecular (within TAZ1, HIF-1$\alpha$, and CITED2) and the inter-molecular (between TAZ1 and HIF-1$\alpha$, between TAZ1 and CITED2) contacts. 
The right panel (B) is the first 400 ns.
}
\label{fig:conver}
\end{figure}

For thermodynamic simulations (binding and unbinding for multiple times),
REMD simulations and long-time MD simulations were performed to overcome the energy barriers between bound and unbound states.
We define that a native contact is formed if the C$\alpha$-C$\alpha$ distance between any given native atom pair is within 1.2 times of its native distance.
Then the profiles of free energy curve or surface can be obtained by using WHAM algorithm \cite{kumar1992weighted,kumar1995multidimensional}.
The REMD simulations have been tested to be converged that the fraction of all the native contacts in the ternary system (Q) becomes equilibrated and stable after about 150 ns of each replica (see \autoref{fig:conver}).

For kinetic simulations, 200 individual MD runs started with varying configurations and velocities were performed on different processes respectively:
direct binding (both unbound state to TAZ1-HIF-1$\alpha$ or TAZ1-CITED2 state)
and replacement (CITED2 binding to TAZ1 by replacing HIF-1$\alpha$ and HIF-1$\alpha$ binding to TAZ1 by replacing HIF-1$\alpha$).

\subsection{$\phi$ value of binding}
The calculation of $\phi$ value is referred to the previous simulation papers \cite{levy2005survey,clementi2000topological}.
The $\phi_{ij}$ for each inter-molecular native contact pair between residue $i$ and $j$ was computed from the probability of formation, $P_{ij}$:
\begin{equation}
\phi_{ij} = \frac{\Delta \Delta F ^{TS-U}}{\Delta \Delta F ^{B-U}}\approx \frac{P_{ij}^{TS} - P_{ij}^{U}}{P_{ij}^{B} - P_{ij}^{U}}
\label{eq:phi}
\end{equation}
where $\Delta \Delta F$ is the free energy difference between the wild-type and mutated protein, $P_{ij}$ is the 
probability of formation of contact between $i$ and $j$.
Here, U, TS, B correspond to the unbound state, transition state ($Q_{inter} \sim 0.03-0.1$), and bound state, respectively.
Then, $\phi_i$ value of residue $i$ can be calculated from the average of $\phi_{ij}$ that are involved with residue $i$.

\newpage

\begin{figure}
\centering
\includegraphics[width=0.6\textwidth]{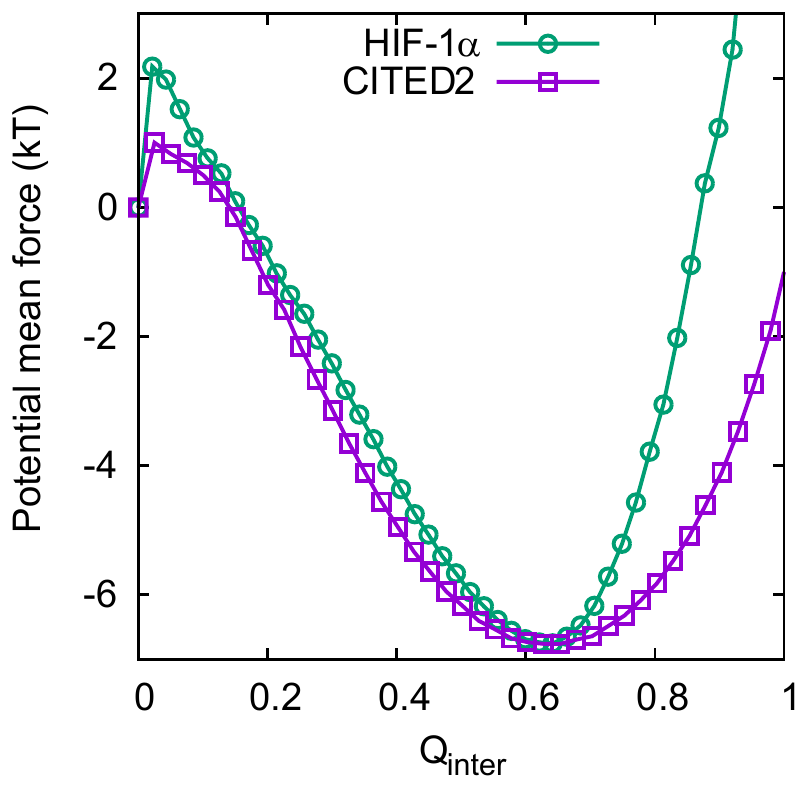}
\caption{
The free energy curves as a function of the fraction of inter-molecular native contacts ($Q_{inter}$) of TAZ1-HIF-1$\alpha$ (green line) and TAZ1-CITED2 (magenta line). 
The free energy unit is kT (k is Boltzmann constant).
}
\label{fig:bi-1d}
\end{figure}

\begin{figure}
\centering
\includegraphics[width=1.0\textwidth]{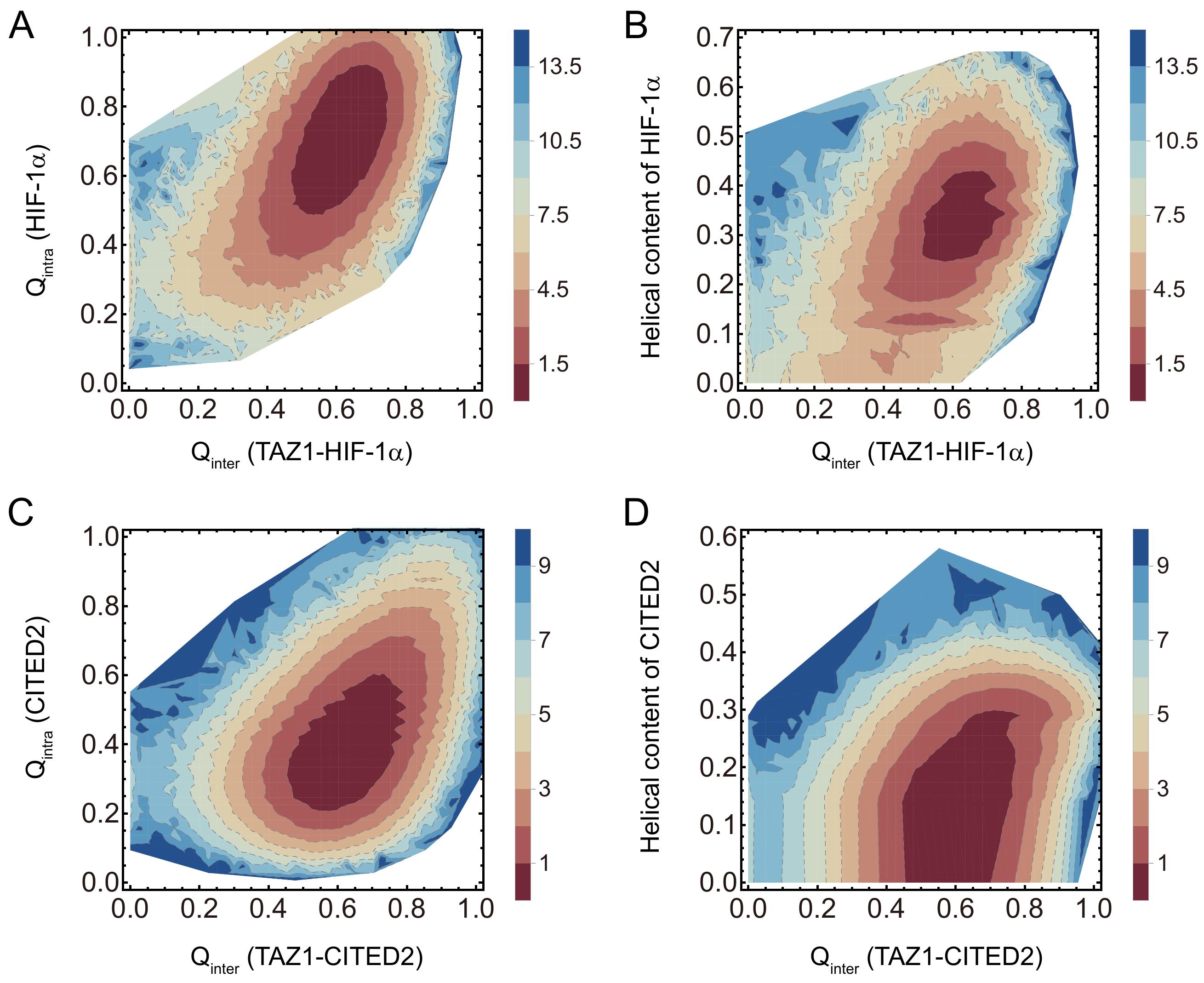}
\caption{
The free energy surfaces projected on (A) both the fraction of inter-molecular native contacts ($Q_{inter}$) of TAZ1-HIF-1$\alpha$ and the fraction of intra-molecular native contacts ($Q_{intra}$) of HIF-1$\alpha$;
(B) both $Q_{inter}$ of TAZ1-HIF-1$\alpha$ and the helical content of HIF-1$\alpha$;
(C) both $Q_{inter}$ of TAZ1-CITED2 and $Q_{intra}$ of CITED2;
(D) both $Q_{inter}$ of TAZ1-CITED2 and the helical content of CITED2.
The free energy unit is kT (k is Boltzmann constant).
}
\label{fig:bi-2d}
\end{figure}

\begin{figure}
\centering
\includegraphics[width=1.0\textwidth]{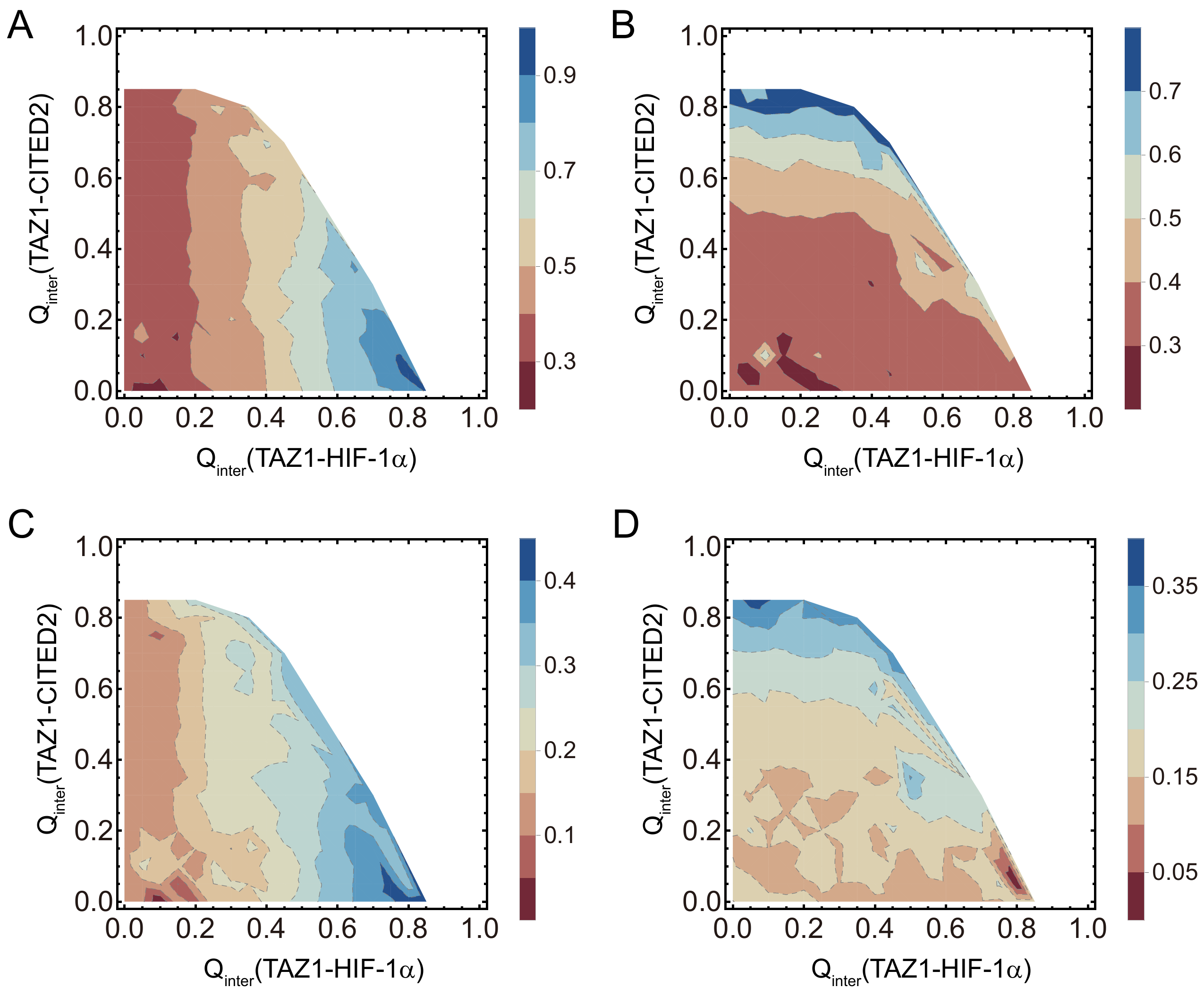}
\caption{
Mean $Q_{intra}$ of HIF-1$\alpha$ (A) and CITED2 (B) as well as mean helical content of HIF-1$\alpha$ (C) and CITED2 (D)
as a function of $Q_{inter}$ (TAZ1-HIF-1$\alpha$) and $Q_{inter}$ (TAZ1-CITED2).
}
\label{fig:mean-H-a}
\end{figure}

\begin{figure}
\centering
\includegraphics[width=1.0\textwidth]{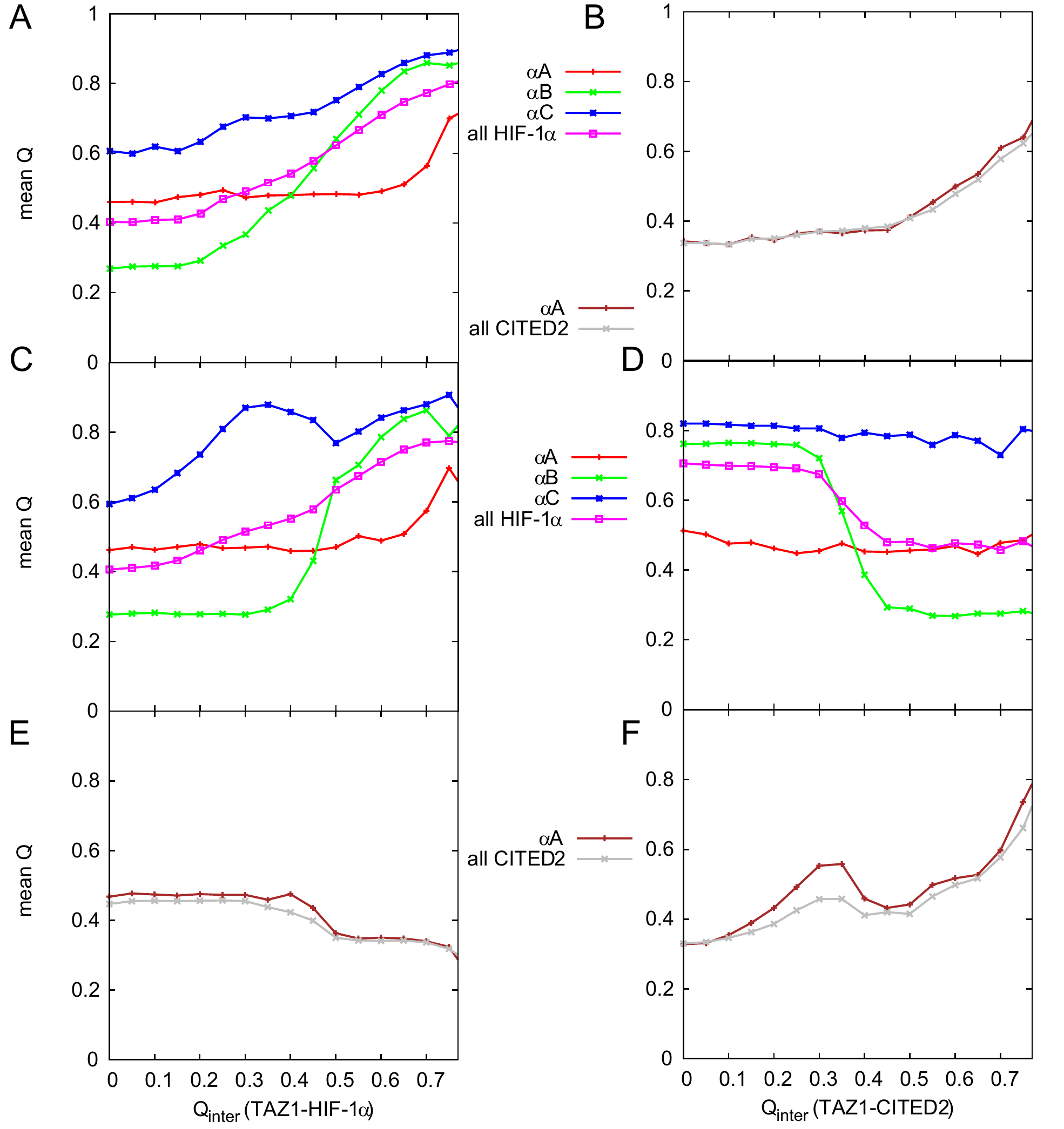}
\caption{
The mean $Q_{intra}$ within different parts of ligand (HIF-1$\alpha$ or CITED2) as a function of binding
in the UH (A), UC (B), CIH (C and E), and HIC (D and F) pathways.
Panels A, C, and D show the $Q_{intra}$ curves of HIF-1$\alpha$; Panels B, E, and F show the $Q_{intra}$ curves of CITED2.
}
\label{fig:distria-all}
\end{figure}

\begin{figure}
\centering
\includegraphics[width=1.0\textwidth]{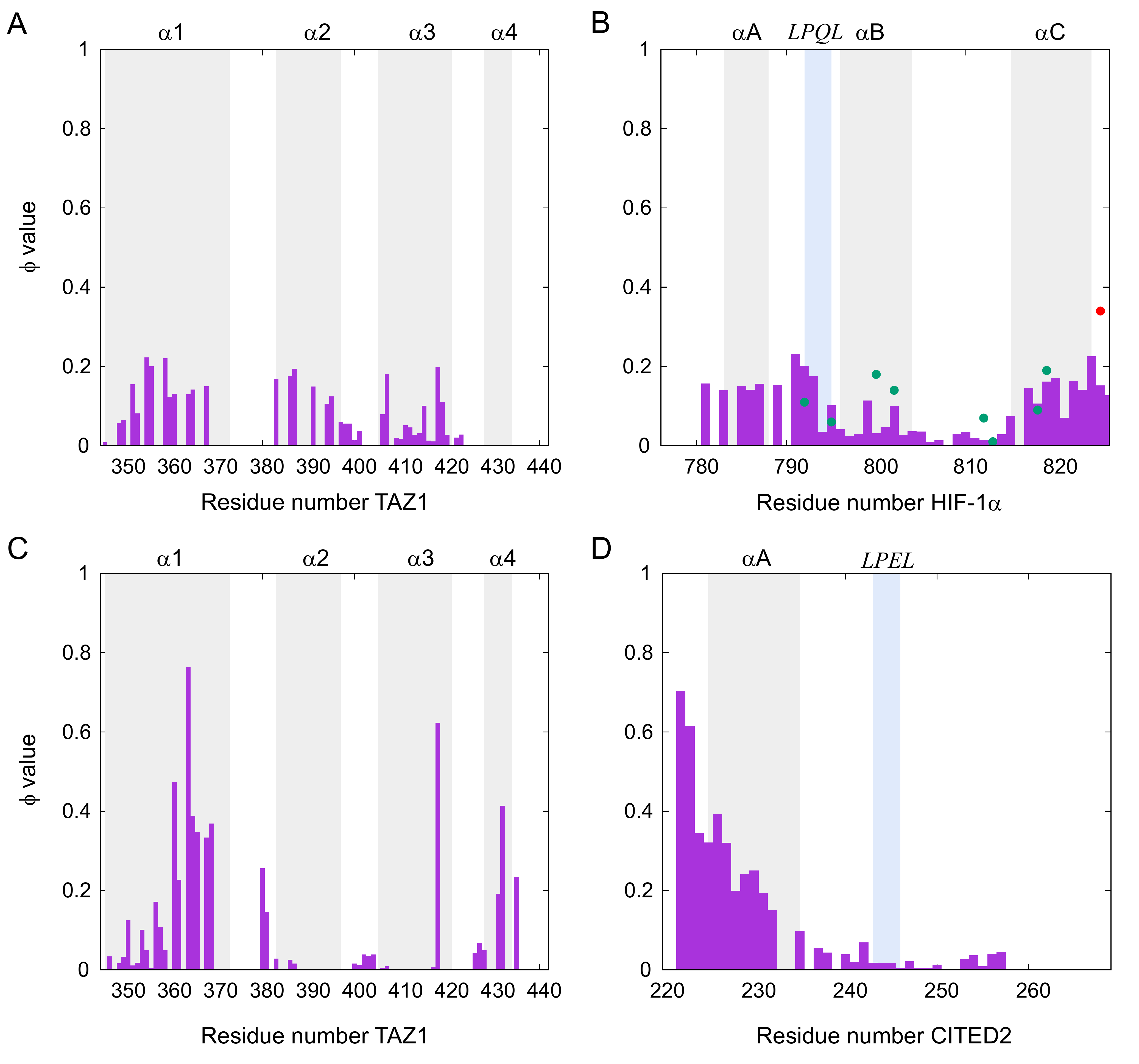}
\caption{
The $\phi$ values of TAZ1 (residue 345 to 439, as well as 3 Zn$^{2+}$) in UH (A) and UC (C) pathways as well as 
the $\phi$ values of HIF-1$\alpha$ (residue 776 to 826) in UH pathway (B) and CITED2 (residue 220 to 269) in UC pathway (D).
The experimental $\phi$ values (in ref\cite{lindstrom2018transition}) are shown in dots.
The secondary structures as well as the LPQL/LPEL motif are labeled.
}
\label{fig:phi1}
\end{figure}

\begin{figure}
\centering
\includegraphics[width=1.0\textwidth]{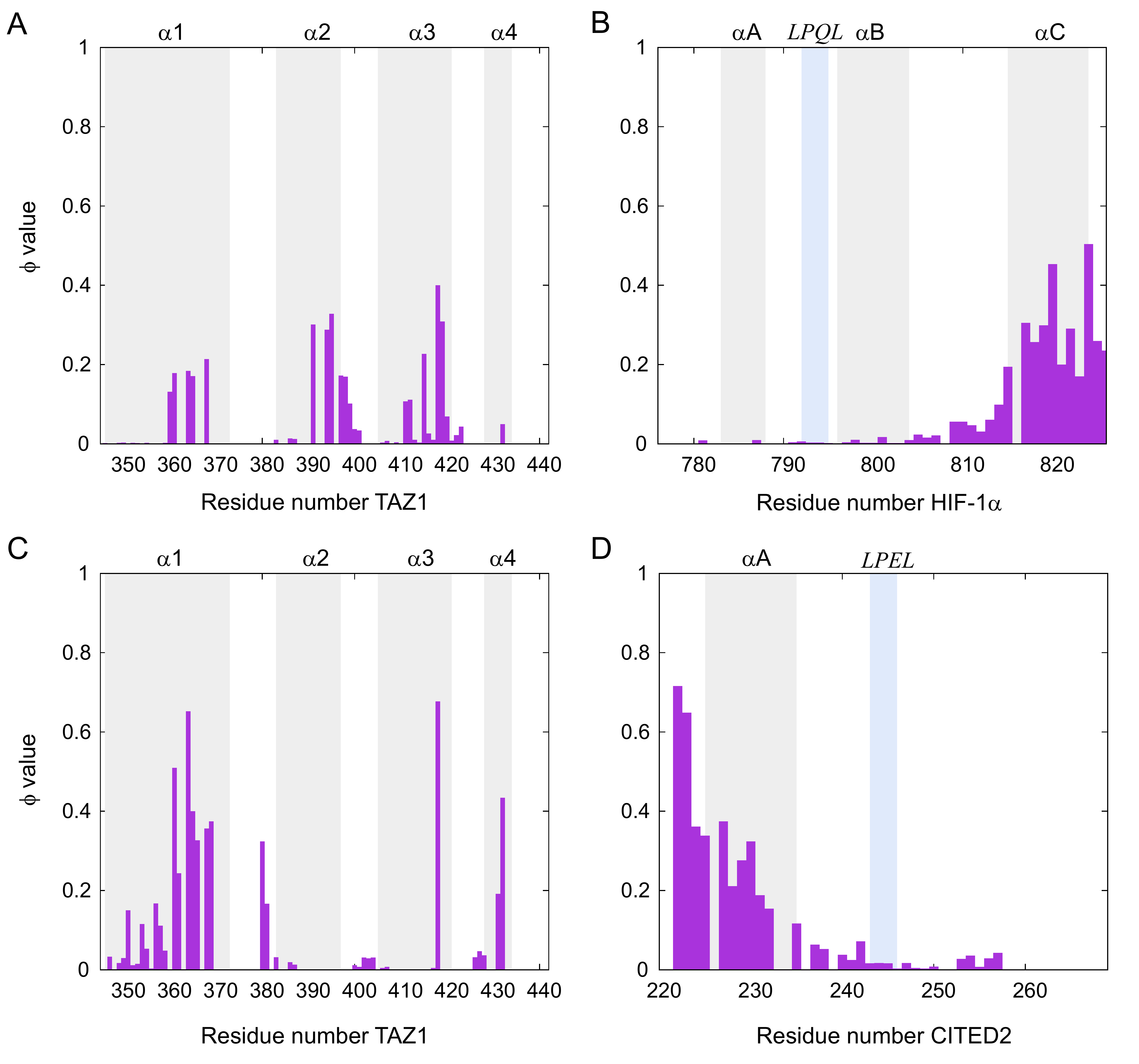}
\caption{
The $\phi$ values of TAZ1 (residue 345 to 439, as well as 3 Zn$^{2+}$) in CIH (A) and HIC (C) pathways as well as 
the $\phi$ values of HIF-1$\alpha$ (residue 776 to 826) in CIH pathway (B) and CITED2 (residue 220 to 269) in HIC pathway (D).
The secondary structures as well as the LPQL/LPEL motif are labeled.
}
\label{fig:phi2}
\end{figure}

\newpage

\bibliographystyle{rsc}
\bibliography{TAZ1_abb}